\documentclass[conference]{IEEEtran}
\IEEEoverridecommandlockouts

\usepackage{algorithmic}
\usepackage{graphicx}
\usepackage{textcomp}
\usepackage[dvipsnames]{xcolor}
\def\BibTeX{{\rm B\kern-.05em{\sc i\kern-.025em b}\kern-.08em
    T\kern-.1667em\lower.7ex\hbox{E}\kern-.125emX}}

\usepackage{mathrsfs,mathdots,mathtools,amssymb,centernot,accents}
\usepackage{nicefrac}
\usepackage{stackengine} 

\usepackage{hyperref}  
\hypersetup{colorlinks,
            linkcolor=blue,
            citecolor=blue,
            urlcolor=black,
            anchorcolor=blue,
            filecolor=blue,
            linktocpage,
            plainpages=false,
            breaklinks=true}
\usepackage[nameinlink]{cleveref}

\usepackage{subcaption}
\usepackage[inline]{enumitem}
\setlist[enumerate]{
  leftmargin=*, 
  label=(\roman*)
}

\usepackage{bm}	
\usepackage{multicol}
\usepackage{dirtytalk}
\usepackage{color, soul}
\usepackage{lipsum}
\usepackage{booktabs}

\usepackage[style=ieee,maxnames=2, minnames=1, maxbibnames=99,citestyle=numeric-comp]{biblatex}

\newcommand{\bE}{\ensuremath{\mathbb{E}}}

\newcommand{\bP}{\ensuremath{\mathbb{P}}}

\newcommand{\bR}{\ensuremath{\mathbb{R}}}

\newcommand{\cA}{\ensuremath{\mathcal{A}}}
\newcommand{\cB}{\ensuremath{\mathcal{B}}}

\newcommand{\cE}{\ensuremath{\mathcal{E}}}

\newcommand{\cG}{\ensuremath{\mathcal{G}}}

\newcommand{\cI}{\ensuremath{\mathcal{I}}}

\newcommand{\cK}{\ensuremath{\mathcal{K}}}

\newcommand{\cN}{\ensuremath{\mathcal{N}}}

\newcommand{\cW}{\ensuremath{\mathcal{W}}}
\newcommand{\cX}{\ensuremath{\mathcal{X}}}
\newcommand{\cY}{\ensuremath{\mathcal{Y}}}
\newcommand{\cZ}{\ensuremath{\mathcal{Z}}}

\newcommand{\dinf}{\ensuremath{D_\infty}}
\newcommand{\ind}{\ensuremath{\mathbf{1}}}
\newcommand{\Ren}{R\'enyi }
\newcommand{\supp}{\ensuremath{\operatorname{supp}}}

\DeclarePairedDelimiter\abs{\lvert}{\rvert}%

\DeclareMathOperator*{\esssup}{ess\,sup}

\newtheorem{theorem}{Theorem}
\newtheorem{definition}[theorem]{Definition}
\newtheorem{cor}{Corollary}

\newtheorem{lemma}{Lemma}

\crefdefaultlabelformat{#2\textbf{#1}#3}
\Crefname{definition}{\textbf{Definition}}{\textbf{Definitions}}
\Crefname{lemma}{\textbf{Lemma}}{\textbf{Lemmas}}
\Crefname{theorem}{\textbf{Theorem}}{\textbf{Theorems}}

\allowdisplaybreaks
    
\begin{document}

\title{On the Information Leakage Envelope of the \\ Gaussian Mechanism
\thanks{This work was supported by the Swedish Research Council (VR) under grant 2024-06615.}
}

\author{\IEEEauthorblockN{Sara Saeidian}
\IEEEauthorblockA{KTH Royal Institute of Technology, 100 44 Stockholm, Sweden, saeidian@kth.se}
  \IEEEauthorblockA{Inria Saclay, 91120 Palaiseau, France}
}

\maketitle

\begin{abstract}
We study the \emph{pointwise maximal leakage} (PML) envelope of the Gaussian mechanism, which characterizes the smallest information leakage bound that holds with high probability under arbitrary post-processing. For the Gaussian mechanism with a Gaussian secret, we derive a closed-form expression for the \emph{deterministic} PML envelope for sufficiently small failure probabilities. We then extend this result to general unbounded secrets by identifying a sufficient condition under which the envelope coincides with the Gaussian case. In particular, we show that strongly log-concave priors satisfy this condition via an application of the Brascamp–Lieb inequality.
\end{abstract}


\section{Introduction}
\label{sec:intro}
Effective data protection requires formal frameworks that enable the use of analytical arguments for quantifying privacy. 
Let $X$ be a random variable representing a secret, and $Y$ the output of a \emph{privacy mechanism} (a conditional probability kernel) $P_{Y \mid X}$ with input $X$. From an information-theoretic perspective, a natural measure of dependence between $X$ and $Y$ is the \emph{information density}
\begin{equation*}
i_{P_{XY}} = \log \frac{d P_{XY}}{d (P_X \times P_Y)}.
\end{equation*}
Different privacy definitions are obtained by imposing different constraints on the information density, many of which admit strong operational interpretations. In particular, it has been shown that upper bounds on the information density limit the inference gain of opportunistic adversaries seeking to maximize arbitrary gain functions~\cite{saeidian2023pointwise_it, saeidian2023pointwise_isit}, while lower bounds restrict the inference gain of risk-averse adversaries aiming to minimize arbitrary risk functions~\cite{saeidianInformationDensityBounds2026,kurriMaximal2024}. More generally, one may seek to control both upper and lower bounds simultaneously~\cite{6483382,jiang2020LIP,jiang2021LIPcontextaware,zarrabian2022asymmetric,zarrabian2023lift}.

While uniform bounds on the information density provide strong and operationally meaningful privacy guarantees, they may be too restrictive in practice. For instance, consider the canonical setting in which independent, zero-mean Gaussian noise is added to a zero-mean Gaussian secret. In this case, the information density between $X$ and $Y$ is
\begin{equation*}
    i_{P_{XY}}(x;y) = \frac{1}{2}\log \left(\frac{\sigma_X^2+\sigma_N^2}{\sigma_N^2}\right) - \frac{(y-x)^2}{2\sigma_N^2} + \frac{y^2}{2(\sigma_X^2+\sigma_N^2)},
\end{equation*}
for $x,y \in \bR$, where $\sigma_X^2$ denotes the variance of $X$ and $\sigma_N^2$ denotes the noise variance. Here, the information density is clearly unbounded. Hence, in many common settings, it may be desirable to define privacy guarantees by putting constraints on the tails of the information density.

In this paper, we consider one such approach using the framework of \emph{pointwise maximal leakage} (PML)~\cite{saeidian2023pointwise_it, saeidian2023pointwise_isit}. PML quantifies the information leaking about $X$ due to releasing $Y$, and can be expressed as
\begin{equation*}
    \ell_{P_{XY}}(X \to Y) = \esssup_{P_X} i_{P_{XY}}(X;Y).
\end{equation*}
The joint distribution $P_{XY}$ is said to satisfy $\varepsilon$-PML if $\ell_{P_{XY}}(X \to Y) \leq \varepsilon$ almost surely for some $\varepsilon > 0$. When no finite $\varepsilon$ satisfies this condition, a natural relaxation is to require the bound to hold with high probability. However, as observed in~\cite{saeidian2023pointwise_it}, simply bounding the tail of $\ell_{P_{XY}}(X \to Y)$ does not yield a robust privacy definition, since the resulting guarantee is not \emph{closed under post-processing}. Specifically, there exist cases where the PML of a $Z$ satisfying the Markov chain $X - Y - Z$ has a heavier tail than that of $Y$, i.e., 
\begin{equation*}
    \bP \{\ell_{P_{XZ}}(X \to Z) > \varepsilon \} > \bP \{\ell_{P_{XY}}(X \to Y) > \varepsilon \},
\end{equation*}
for fixed $\varepsilon > 0$. It is worth emphasizing that closedness under post-processing is often regarded as an axiom for privacy definitions, motivated by the data-processing inequality~\cite{kifer2012axiomatic}.

In response to this observation, our recent work~\cite{csf_envelope} introduced the notion of the \emph{PML envelope}. Let
\begin{equation*}
    \delta_c(\varepsilon) \coloneqq \sup_{Z: X - Y - Z} \bP\{\ell_{P_{XZ}}(X \to Z) > \varepsilon\}, \quad \varepsilon >0,
\end{equation*}
be the largest \emph{failure probability} over all post-processed versions of $Y$. The PML envelope is then defined as
\begin{equation*}
    \varepsilon_c(\delta) \coloneqq \inf\{\varepsilon > 0 : \delta_c(\varepsilon) \le \delta\}, \quad \delta \in (0,1).
\end{equation*}
Intuitively, $\varepsilon_c(\delta)$ is the tightest upper bound on the information leakage that holds with probability at least $1 - \delta$ under any downstream transformation of $Y$.

In~\cite{csf_envelope}, we characterized the PML envelope in multiple settings where both $X$ and $Y$ had finite alphabets. When all underlying probability spaces are finite, PML admits a uniform upper bound and therefore satisfies $\varepsilon$-PML for some finite $\varepsilon>0$. The real strength of the PML envelope, however, emerges in infinite settings, where no uniform bound exists. In this work, we build on~\cite{csf_envelope} by characterizing the PML envelope of the \emph{Gaussian mechanism}, namely, the mechanism that releases a real-valued secret~$X$ perturbed by additive, independent Gaussian noise. This mechanism is a compelling object of study for several reasons. First, Gaussian mechanisms are central in the framework of \emph{differential privacy} (DP) \cite{dworkAlgorithmicFoundationsDifferential2014}, and as such, are widely used in modern data-processing pipelines. Second, Gaussian noise arises naturally in many systems due to the central limit phenomenon. Third, the Gaussian distribution is analytically convenient, with many well-understood theoretical properties.

\textbf{Contributions.} Here, we focus on characterizing the \emph{deterministic PML envelope} for the Gaussian mechanism, denoted by $\varepsilon_d(\delta)$. This quantity captures the worst-case information leakage under arbitrary deterministic post-processings of $Y$ with failure probability at most $\delta$. By definition, $\varepsilon_d$ provides an achievable lower bound on $\varepsilon_c$. 

Our main contributions are summarized as follows:
\begin{itemize}[leftmargin=*]
    \item For the Gaussian mechanism with a Gaussian secret, we derive a closed form for $\varepsilon_d(\delta)$ for sufficiently small $\delta$ (Theorem~\ref{thm:Gauss_mech_Gauss_secret_envelope}). 

    \item For the Gaussian mechanism with a general unbounded secret, we establish a sufficient condition under which the deterministic PML envelope coincides with the Gaussian case for sufficiently small $\delta$ (Theorem~\ref{thm:Gauss_mech_envelope}). 
    

    \item We use the Brascamp-Lieb inequality~\cite{brascampExtensionsBrunnMinkowskiPrekopaLeindler1976a} to show that \emph{strongly log-concave} distributions \cite{saumardLogconcavityStrongLogconcavity2014} satisfy the condition identified in Theorem~\ref{thm:Gauss_mech_envelope}.

\end{itemize}

\textbf{Related works.} The problem of robustness under post-processing for privacy guarantees based on tail probabilities has also been identified in the DP literature. In particular, \textcite{meiser2018approximate} showed that probabilistic relaxations of DP are not closed under post-processing. This observation has motivated a range of definitions that characterize guarantees in terms of the distribution or moments of the privacy loss. Notable examples include \emph{concentrated DP}~\cite{bunConcentratedDifferentialPrivacy2016a}, \emph{\Ren DP}~\cite{mironovRenyiDifferentialPrivacy2017a}, and \emph{Gaussian DP}~\cite{dongGaussianDifferentialPrivacy2022}. The closest analogue to the PML envelope in the DP literature is the \emph{privacy profile}, which uses a tradeoff curve between the privacy parameters~\cite{ballePrivacyProfilesAmplification2020a}. The underlying parameterization, however, differs from our work: privacy profiles are formulated in terms of the additive parameter $\delta$ from \emph{approximate DP}~\cite{dworkAlgorithmicFoundationsDifferential2014}, whereas here, $\delta$ quantifies the failure probability of the leakage bound (see~\cite{csf_envelope} for details).




\section{Background}
\label{sec:background}
\textbf{Notation.} 
For a random variable $X$, $P_X$ denotes its distribution. When $P_X \ll \lambda$, with $\lambda$ the Lebesgue measure on $\mathbb{R}$, we write $f_X = \frac{dP_X}{d\lambda}$ to denote its density. We use analogous notation for joint and conditional distributions. For a measurable function $h(X)$, the notation $\esssup_{P_X} h(X)$ denotes the essential supremum with respect to the distribution $P_X$. For a measurable set $\cE$, $\ind_{\cE}$ denotes its indicator function.

Consider an adversary whose goal is to maximize a non-negative gain function $g$, where $g$ can represent a wide range of privacy attacks such as membership or attribute inference. To quantify the information leakage associated with a single released outcome $y \in \cY$, PML compares the expected value of $g$ after observing $y$ to its expected value prior to observing $y$ in a ratio. Then, to obtain a robust measure, this ratio of posterior-to-prior expected gain is maximized over all possible non-negative gain functions. 

\begin{definition}[{\cite{saeidian2023pointwise_isit}}]
\label{def:general_pml}
Given a joint distribution $P_{XY}$, the pointwise maximal leakage from $X$ to $y \in \cY$ is defined as 
\begin{equation*}
    \ell_{P_{XY}}(X \to y) \coloneqq \log \sup_{\substack{g \in \cG}} \frac{\sup\limits_{P_{\hat W \mid Y}} \bE \left[g(X,\hat W) \mid Y=y \right]}{\sup\limits_{w' \in \cW} \, \bE[g(X,w')]},
\end{equation*}
where $\cG$ denotes the set of all non-negative measurable gain functions $g: \cX \times \cW \to \bR_+$ satisfying $\sup\limits_{w'} \, \bE[g(X,w')] < \infty$.
\end{definition}

In this work, both $X$ and $Y$ are real-valued and satisfy
$P_X \ll \lambda$ and $P_{Y \mid X = x} \ll \lambda$ for all $x \in \mathbb{R}$. Under these assumptions, PML admits the simplified representation
\begin{align*}
    &\ell_{P_{XY}}(X \to y) =  \dinf (P_{X \mid Y=y} \Vert P_{X})\\
    &= \esssup_{P_X} \; i_{P_{XY}}(X; y) = \esssup_{P_X} \; \log \frac{f_{Y \mid X}(y \mid X)}{f_Y(y)},
\end{align*}
where $\dinf(\cdot \Vert \cdot)$ denotes the \Ren divergence of order infinity~\cite{renyi1961measures}, $P_{X \mid Y}$ denotes the posterior distribution and $i_{P_{XY}}$ denotes the information density of $P_{XY}$ \cite{saeidian2023pointwise_isit}. When the information density is continuous in $x$ (as is the case for the Gaussian mechanism), the essential supremum coincides with the supremum over the support of $X$, denoted by $\supp(P_X)$. Moreover, when the underlying joint distribution is clear from context, we omit the subscript $P_{XY}$ and write $i(x;y)$ and $\ell(X \to y)$, or simply $\ell(y)$, for $x,y \in \mathbb{R}$.

Throughout the paper, we will also consider the information leakage associated with measurable events $\cE \subseteq \mathcal{Y}$. This is simply a shorthand for the PML incurred by revealing that $Y \in \cE$. Specifically, letting $Z = \ind_{\cE}(Y)$, we define
\begin{equation*}
    \ell_{P_{XY}}(X \to \cE) \coloneqq \ell_{P_{XZ}}(X \to 1) = \esssup_{P_X} \; \log \frac{P_{Y \mid X}(\cE \!\mid \!X)}{P_Y(\cE)}.
\end{equation*}
This notation is convenient, as computing the (deterministic) PML envelope requires evaluating the leakage associated with post-processed versions of $Y$. In particular, for deterministic post-processings, each outcome $z \in \cZ$ corresponds to an event $\cE_z = \{y \in \cY : h(y) = z\}$ for some measurable function $h$. 

The distribution $P_{XY}$ is said to satisfy $\varepsilon$-PML with $\varepsilon > 0$ if $\ell(X \to Y) \leq \varepsilon$ almost surely. In \cite{saeidian2023pointwise_it}, Saeidian et al.\ introduced a relaxation of $\varepsilon$-PML by incorporating an additional parameter $\delta \in (0,1)$. Specifically, a distribution $P_{XY}$ is said to satisfy $(\varepsilon, \delta)$-PML if $P_Y \{ \ell(X \to Y) > \varepsilon \} \leq \delta$. However, this definition is not closed under post-processing: an adversary may apply a channel $P_{Z \mid Y}$ to obtain $Z$ such that
\begin{equation*}
P_Z \{ \ell(X \to Z) > \varepsilon \} > P_Y \{ \ell(X \to Y) > \varepsilon \}.
\end{equation*}

To achieve post-processing robustness, we therefore consider the worst-case probability of failure over all possible downstream transformations of the released variable. Specifically, for a given joint distribution $P_{XY}$ and $\varepsilon > 0$, we define the \emph{closed failure probability}
\begin{equation*}
    \delta_c(\varepsilon) \coloneqq \sup_{Z: X - Y - Z} \bP\{\ell(X \to Z) > \varepsilon\}.
\end{equation*}

Rather than fixing $\varepsilon$ and bounding the worst-case failure probability, we can equivalently fix a target failure probability $\delta \in (0,1)$ and ask for the smallest leakage level that survives arbitrary post-processing. This leads to the definition of the \emph{PML envelope}
\begin{equation}
\label{eq:pml_envelope}
    \varepsilon_c(\delta) \coloneqq \inf\{\varepsilon > 0 : \delta_c(\varepsilon) \le \delta\}.
\end{equation}
In \cite{csf_envelope}, we showed that the PML envelope admits several equivalent characterizations. The form most convenient for our present analysis is
\begin{equation}
\label{eq:envelope}
    \varepsilon_c(\delta) = \sup_{Z : X - Y - Z} \; \bar \varepsilon_Z(\delta),
\end{equation}
where
\begin{equation*}
    \bar \varepsilon_Z(\delta) \coloneqq \sup \{t \geq 0 : P_Z \{\ell(Z) > t\} \geq \delta \}, 
\end{equation*}
is called the \emph{PML $\delta$-quantile} of $Z$. If $Z$ is a finite random variable, this quantity admits the more explicit representation
\begin{equation*}
   \bar \varepsilon_Z(\delta)
   = \max_{\substack{\cA \subset \cZ \\ P_Z(\cA) \geq \delta}}
   \; \min_{z \in \cA \cap \supp(P_Z)} \; \ell(z).
\end{equation*}
This representation will be used repeatedly in the proofs.

\section{Deterministic PML Envelope of the \\ Gaussian Mechanism}
\label{sec:gaussian}
Here, we focus on a restricted version of the PML envelope in which the post-processing is required to be deterministic. Specifically, we define
\begin{equation*}
    \varepsilon_d(\delta) := \sup_{Z = h(Y)} \; \bar{\varepsilon}_Z(\delta),
\end{equation*}
where the supremum is taken over all measurable functions $h$. By construction, $\varepsilon_d(\delta) \le \varepsilon_c(\delta)$ for all
$\delta \in (0,1)$.

We characterize $\varepsilon_d$ for the canonical example of the \emph{Gaussian mechanism}, i.e., the system $Y = X + N$, where $N \sim \cN(0, \sigma_N^2)$ is zero-mean Gaussian noise and independent of $X$. Without loss of generality, we assume that $\bE [X] = 0$. 

\subsection{Properties of the Gaussian Mechanism} 
The Gaussian mechanism has been extensively studied in information theory (e.g.,~\cite{guoMutualInformationMinimum2005, guoEstimationGaussianNoise2011, dytsoConditionalMeanEstimation2023}) as well as in mathematics and physics through its connection to the heat equation~\cite{widderHeatEquation2010}. In this section, we state some properties of this mechanism that are central to the analysis of the PML envelope.

Let $\Phi : \bR \to (0,1)$ denote the CDF of the standard Gaussian distribution, and let $\varphi : \bR \to \bR_+$ denote its probability density function (pdf). We begin by observing that the marginal density of $Y$ is 
\begin{equation*}
    f_Y(y) = \frac{1}{\sigma_N} \int_{\bR} f_X(x) \, \varphi\left(\frac{y - x}{\sigma_N}\right) dx, \quad y \in \bR,
\end{equation*}
which is the convolution of $f_X$ with a Gaussian kernel. It follows from standard results (e.g.,~\cite[Remark 4.1]{yanLectureNotesPartial}) that $f_Y$ is a real analytic function on $\bR$. In particular, $f_Y$ is continuous and infinitely differentiable, and all of its level sets, i.e., sets of the form $\{y \in \bR : f_Y(y) = c\}$ for $c \geq 0$, have Lebesgue measure zero~\cite[Sec.~4.1]{krantzPrimerRealAnalytic2002}. Furthermore, for each fixed $x \in \bR$, the information density $i(x; y)$ is real analytic in $y$, since both $f_{Y \mid X=x}$ and $f_Y$ are real analytic functions. Importantly, both of these properties hold for arbitrary distributions on $X$.

Below, we state a lemma that is a key technical component of our analysis. Computing $\varepsilon_d$ requires evaluating the information leakage associated with various events and in countable probability spaces, this can be done without difficulty. However, for more general settings such as the Gaussian mechanism, it is not feasible to compute the leakage of arbitrary measurable sets, which may be highly irregular (e.g., the Cantor set). Therefore, our first step is to show that instead of all measurable sets, we can restrict our attention to \emph{intervals}, often in the tails of $Y$.

\begin{lemma}
\label{lemma:tail_worse}
Consider the Gaussian mechanism $Y = X + N$, where $N \sim \cN (0, \sigma_N^2)$ and $X$ and $N$ are independent. Suppose $X$ is unbounded, i.e., $\supp(P_X) = \bR$.  
\begin{enumerate}
\item Let $\delta \in (0,1)$ and suppose $t_L, t_R \in \bR$ satisfy
\begin{equation*}
    P_Y (-\infty, t_L) = P_Y (t_R, \infty) = \delta. 
\end{equation*} 
Then, it holds that
\begin{multline*}
\label{eq:worst_set_unbounded}
    \sup_{\cE \subset \bR : P_Y(\cE) \geq \delta} \; \ell (X \to \cE) \leq \ell(X \to (-\infty, t_L))\\
    = \ell(X \to (t_R, \infty)) = \log \frac{1}{\delta}. 
\end{multline*}

\item Let $(a,b) \subset \bR$ be an interval with $-\infty < a < b < \infty $, and let $0 < \delta < P_Y (a,b)$. Then, there exists an interval $\cA^* \subset (a,b)$ such that 
\begin{equation*}
\label{eq:worst_set_bounded}
    \ell (X \to \cA^*) = \sup_{\cE \subset (a,b) : P_Y(\cE) \geq \delta} \; \ell (X \to \cE). 
\end{equation*}

\item Suppose there exists $M >0$ such that $f_Y$ is (strictly) increasing on $(-\infty, -M)$ and (strictly) decreasing on $(M, \infty)$. Then, there exist $a_0, b_0 \in \bR$ and $\delta_0 \in (0,1)$ such that for all $a\leq a_0$, $b \geq b_0$ and $0 < \delta \leq \delta_0$, we have 
\begin{multline*}
\label{eq:worst_set_tails}
    \sup_{\cE \subset (a,b) : P_Y(\cE) \geq \delta} \; \ell (X \to \cE)\\
    =\max\Big\{ \ell (X \to (a,a_\delta)), \ell (X \to (b_\delta,b)) \Big\},
\end{multline*}
where $a_\delta$ and $b_\delta$ satisfy 
\begin{equation*}
    P_Y(a,a_\delta) = P_Y (b_\delta,b) = \delta.
\end{equation*}
\end{enumerate}
\end{lemma}
\begin{IEEEproof}
See Appendix~\ref{ssec:lemma_tail_worse_proof}. 
\end{IEEEproof}

\subsection{Gaussian Secret}
Here, we assume that $X$ is Gaussian and compute $\varepsilon_d$ of the Gaussian mechanism for this particular secret. By Lemma~\ref{lemma:tail_worse}, the events with the largest leakage are always intervals. We therefore begin by computing the information leakage of intervals. 

\begin{lemma}
\label{lemma:gaussian_interval_leakage}
Let $X \sim \cN(0, \sigma_X^2)$ and $N \sim \cN(0,\sigma_N^2)$. The Gaussian noise mechanism $Y = X + N$ has information leakage 
\begin{equation}
\label{eq:event_pml_Gaussian_1}
    \ell(X \to (a,b)) = \log \; \frac{2\Phi(\frac{b-a}{2 \sigma_N}) -1}{\Phi(\frac{b}{\sigma_Y}) - \Phi(\frac{a}{\sigma_Y})}, \quad (a,b) \subset \bR,
\end{equation}
where $\sigma_Y^2 = \sigma_X^2 + \sigma_N^2$ denotes the variance of $Y$.
\end{lemma}
\begin{IEEEproof}
See Appendix~\ref{ssec:gaussian_interval_leakage_proof}.
\end{IEEEproof}

The next result gives a closed-form expression for $\varepsilon_d$ assuming a Gaussian secret.
\begin{theorem}
\label{thm:Gauss_mech_Gauss_secret_envelope}
Let $Y = X + N$ with $X \sim \cN(0, \sigma_X^2)$ and $N \sim \cN (0, \sigma_N^2)$. If $\sigma_X^2 \leq 3 \sigma_N^2$, then 
\begin{equation*}
    \varepsilon_d(\delta) = \log \frac{2}{\delta},
\end{equation*}
for all $\delta \in (0,0.5)$. 
\end{theorem}
\begin{IEEEproof}
Fix $\delta \in (0,0.5)$. Let $t_\delta > 0$ be such that $P_Y \{y \in \bR : \abs{y} > t_\delta\} = \delta$, and observe that $P_Y (t_\delta, \infty) = P_Y (-\infty, -t_\delta) = \frac{\delta}{2}$. We call $C_\delta = \{y \in \bR : \abs{y} \leq t_\delta\}$ the \say{core} set.

Our goal is to bound $\bar \varepsilon_{h(Y)} (\delta)$ for an arbitrary measurable function $h : \cY \to \cZ$. First, we will assume that $h$ is a simple function, that is, $\cZ$ is a finite set. We will show that in this case, $\bar \varepsilon_{h(Y)} (\delta) \leq \log \frac{2}{\delta}$. Afterwards, we will extend the argument from simple functions to arbitrary measurable functions.  

Let $Z_m = h_m(Y)$, where $h_m$ is a simple function with $m+1$ outcomes and $\cZ_m = \cG \cup \cB_m$, where $\cG$ is the \say{good} set with probability $P_Z(\cG) = 1-\delta$, $\cB_m$ is the \say{bad} set with probability $P_Z(\cB_m) = \delta$, and $\max_{z \in \cG} \ell (z) \leq \min_{z \in \cB_m} \ell (z)$. Recall that our goal is to calculate 
\begin{equation*}
    \bar \varepsilon_{Z_m} (\delta) = \max_{\substack{\cA \subset \cZ_m \\ P_Z(\cA) \geq \delta}} \; \min_{z \in \cA} \; \ell(z) = \min_{z \in \cB_m} \; \ell (z). 
\end{equation*}
Without loss of generality, we assume that $\abs{\cG} = 1$ regardless of $m$. If $\cG$ has more elements, we can simply map them all to a single symbol without affecting the value of $\bar \varepsilon_{Z_m} (\delta)$. 

First, consider an arbitrary $Z_2 = h_2(Y)$ so that $\cB_2$ contains two elements, corresponding to two (disjoint) events $\cE_1, \cE_2 \subset \bR$ with $\delta_1 = P_Y(\cE_1)$,  $\delta_2 = P_Y(\cE_2)$ and $\delta_1 + \delta_2 = \delta$. By \cite[Lemma 1(i)]{csf_envelope}, we have  
\begin{align*}
    \bar \varepsilon_{Z_2} (\delta) &= \min_{z \in \cB_2} \; \ell (z) =  \min \Big\{\ell(X \to \cE_1), \ell(X \to \cE_2) \Big\}\\
    &\leq \max_{\substack{\delta_1, \delta_2 > 0 :\\ \delta_1 + \delta_2 = \delta}} \; \min \Big\{\log \frac{1}{\delta_1}, \log \frac{1}{\delta_2}\Big\} = \log \frac{2}{\delta},   
\end{align*}
with equality if $\cE_1 = (-\infty, t_\delta)$ and $\cE_2 = (t_\delta, \infty)$. In this case, the core set $C_\delta$ is mapped to a single symbol in $\cG$. Let $Z_2^*$ denote this construction.

Next, we show that when $3 \sigma_N^2 \geq \sigma_X^2$, adding more symbols will not increase the PML $\delta$-quantile. Let $Z_3 = h_3(Y)$, where $\cB_3$ has three elements, corresponding to three disjoint events $\cE_1, \cE_2, \cE_3 \subset \bR$. Let $\delta_j = P_Y(\cE_j)$ for $j=1,2,3$ and $\delta_1 + \delta_2 + \delta_3 = \delta$. Our goal is to select $\cE_1, \cE_2, \cE_3$ so that they have as large an information leakage as possible, since we want to find the largest value for 
\begin{equation*}
    \bar \varepsilon_{Z_3} (\delta) = \min_{z \in \cB_3} \; \ell (z) =  \min_{i \in \{1,2,3\}} \ell(X \to \cE_i),
\end{equation*}
subject to the constraints on the probabilities of $\cE_1, \cE_2$ and $\cE_3$. Using Lemma~\ref{lemma:tail_worse}(i), the events with the largest information leakage are the tails, so, we take $\cE_1 = (-\infty, t_L)$ and $\cE_2 = (t_R, \infty)$, where $t_L, t_R$ are picked so that $\cE_1, \cE_2$ have the desired probabilities. For $\cE_3$, Lemma~\ref{lemma:tail_worse}(ii) states that we need to pick an interval in $[t_L, t_R]$ with as large a leakage as possible. From Lemma~\ref{lemma:gaussian_interval_leakage}, it is easy to see that this is the interval adjacent to one of the endpoints $t_L$ and $t_R$. This follows from an argument similar to that in Appendices~\ref{ssec:proof_bathtub_1} and~\ref{ssec:proof_bathtub_2}, by applying the bathtub principle (Theorem~\ref{thm:bathtub}) to~\eqref{eq:event_pml_Gaussian_1}. Note that here, we do not use Lemma~\ref{lemma:tail_worse}(iii) but instead rely on Lemma~\ref{lemma:gaussian_interval_leakage}.

Suppose we take $\cE_3 = (t_C, t_R]$. Note that the condition $\delta < 0.5$ ensures that $t_C > 0$. The following lemma, proved in Appendix~\ref{ssec:proof_lemma_interval_leakage_increasing}, states that when $3 \sigma_N^2 \geq \sigma_X^2$, we have $\ell (X \to \cE_3) \leq \ell(X \to \cE_2)$. 

\begin{lemma}
\label{lemma:interval_leakage_increasing}
If $3 \sigma_N^2 \geq \sigma_X^2$, then for all $a >0$, the mapping $b \mapsto \ell(X \to (a,b))$ is strictly increasing and has the limit 
\begin{equation*}
    \lim_{b \to \infty} \ell(X \to (a,b)) = \log \; \frac{1}{P_Y (a, \infty)}. 
\end{equation*}    
\end{lemma}

Since $t_C > 0$, the lemma applies with $a = t_C$, and we get  
\begin{multline*}
    \ell(X \to \cE_3) \leq \ell(X \to (t_C, \infty)) = \ell(X \to \cE_3 \cup \cE_2)\\
    \leq \ell(X \to (t_R, \infty)) = \ell(X \to \cE_2).  
\end{multline*}
Therefore, we have 
\begin{align*}
    \bar \varepsilon_{Z_3} (\delta) &= \min \Big\{\ell(X \to \cE_1), \ell(X \to \cE_2), \ell(X \to \cE_3) \Big\}\\
    &= \min \Big\{\ell(X \to \cE_1), \ell(X \to \cE_3) \Big\}\\
    &\leq \min \Big\{\ell(X \to \cE_1), \ell(X \to \cE_3 \cup \cE_2) \Big\}\\
    &\leq \max_{\substack{\delta_1, \delta_2, \delta_3 > 0 :\\ \delta_1 + \delta_2 + \delta_3 = \delta}} \; \min \Big\{\log \frac{1}{\delta_1}, \log \frac{1}{\delta_2 + \delta_3}\Big\}\\
    &= \log \frac{2}{\delta}. 
\end{align*}
Similarly, it can be shown that taking $\cE_3 = [t_L, t'_C)$ cannot increase the PML $\delta$-quantile beyond $\log \frac{2}{\delta}$. Thus, no $Z_3$ random variable can have a larger PML $\delta$-quantile compared to $Z^*_2$. 

The above argument can be extended to show that no $Z_m$ can exceed the PML $\delta$-quantile of $Z_2^*$. Indeed, the events with the largest information leakage are intervals in the tail of the Gaussian, so the \say{worst} $Z_m$ slices the left and right tails into adjacent intervals. Then, by Lemma~\ref{lemma:interval_leakage_increasing}, collapsing all the right-tail segments into $(t_R,\infty)$ and, symmetrically, all the left-tail segments into $(-\infty,t_L)$ with $P_Y(t_R,\infty) + P_Y(-\infty,t_L) = \delta$ will increase the PML $\delta$-quantile up to $\log \frac{2}{\delta}$. We conclude that 
\begin{equation*}
    \bar{\varepsilon}_{h(Y)}(\delta) \leq \log \frac{2}{\delta},
\end{equation*}
for every simple function $h$.

The final step is to show that arbitrary measurable post-processing cannot beat the bound $\log \frac{2}{\delta}$ either. We state this as a lemma proved in Appendix~\ref{ssec:lemma_simple_to_measurable_proof}.
\begin{lemma}
\label{lemma:simple_to_measurable}
Suppose $\bar\varepsilon_{Z_m}(\delta)\le \log\frac{2}{\delta}$ 
for all $\delta \in (0,0.5)$ and all $m \geq 1$. Then the same bound holds for every measurable function $h :\mathbb{R}\to\mathcal{Z}$ with $\mathcal{Z}$ standard Borel, i.e.,
\begin{equation*}
    \bar\varepsilon_{h(Y)}(\delta)\le \log\frac{2}{\delta}.
\end{equation*}
\end{lemma}
Combining the bound established for simple post-processings with Lemma~\ref{lemma:simple_to_measurable} completes the proof.
\end{IEEEproof}

\subsection{General Unbounded Secret}
We now turn to the setting where the distribution of $X$ is unspecified. We assume that $X$ has full support on $\bR$.

\begin{theorem}
\label{thm:Gauss_mech_envelope}
Let $Y = X + N$ where $N \sim \cN (0, \sigma_N^2)$, $\bE[X] = 0$, $\supp(P_X) = \bR$ and $X$ and $N$ are independent. Suppose there exists $M >0$ such that $f_Y$ is (strictly) increasing on $(-\infty, -M)$ and (strictly) decreasing on $(M, \infty)$. If 
\begin{equation*}
    \operatorname{Var}[X \mid Y=y] \leq 0.75 \sigma_N^2,
\end{equation*}
for all $y \in \bR$, then there exists $\delta_0 \in (0,0.5)$ such that 
\begin{equation*}
    \varepsilon_d(\delta) = \log \frac{2}{\delta},
\end{equation*}
for all $\delta \leq \delta_0$. 
\end{theorem}
\begin{IEEEproof}
See Appendix~\ref{ssec:thm_Gauss_mech_envelope_proof}.
\end{IEEEproof}

Theorem~\ref{thm:Gauss_mech_envelope} states that if the posterior variance of $X$ can be appropriately controlled, then the resulting $\varepsilon_d$ matches that of the Gaussian case (for sufficiently small $\delta$). In what follows, we describe a class of input distributions that satisfy the required variance bound.

\begin{definition}[Log concave and strongly log concave densities]
A density function $f : \bR \to \bR_+$ is said to be log concave if $f = e^{-\theta}$, where $\theta : \bR \to (-\infty, \infty]$ is a convex function. A density function $g$ is said to be $\beta$-strongly log concave if 
\begin{equation*}
    g(x) = h(x) \varphi(\frac{x}{\beta}), \quad x \in \bR,
\end{equation*}
where $h$ is a log concave function, $\varphi$ is the standard Gaussian density, and $\beta >0$. 
\end{definition}
Note that if $g=e^{-\theta}$ with $\theta$ twice differentiable, then $\beta$-strong log-concavity implies that $\theta''(x) \geq \frac{1}{\beta^2}$ for all $x\in\mathbb{R}$.

The following result, known as the \emph{Brascamp–Lieb inequality} \cite{brascampExtensionsBrunnMinkowskiPrekopaLeindler1976a}, provides a Poincaré-type inequality for log-concave probability measures. Specifically, it bounds the variance of a function in terms of its local fluctuations. We state the inequality in the form given in \cite[Prop. 10.1(a)]{saumardLogconcavityStrongLogconcavity2014}.

\begin{theorem}[Brascamp-Lieb inequality]
\label{thm:brascamp_lieb}
Suppose $f_X = \exp(-\theta)$ is a log-concave density on $\bR$, where $\theta$ is strictly convex and twice continuously differentiable. Let $G : \bR \to \bR$ be a continuously differentiable function with $\bE [G^2(X)] < \infty$. Then, 
\begin{equation}
\label{eq:variance_bound_strict_log_concave}
    \operatorname{Var}[G(X)] \leq \bE \Big[ \frac{(G'(X))^2}{\theta''(X)} \Big].
\end{equation}
\end{theorem}
When $f_X$ is $\beta$-strongly log-concave (in addition to satisfying the conditions of Theorem~\ref{thm:brascamp_lieb}), applying \eqref{eq:variance_bound_strict_log_concave} yields the bound
\begin{equation}
\label{eq:variance_bound_strong_log_concave}
    \operatorname{Var}[G(X)] \leq \beta^2 \, \bE \Big[(G'(X))^2 \Big].
\end{equation}

We now apply~\eqref{eq:variance_bound_strong_log_concave} to bound the posterior variance $\operatorname{Var}[X \mid Y = y]$ for arbitrary $y \in \bR$. Suppose $X$ has a log-concave density $f_X = \exp(-\theta)$ satisfying the conditions of Theorem~\ref{thm:brascamp_lieb}. Fix $y \in \bR$. Under the Gaussian mechanism, the joint density has the form
\begin{align*}
    f_{XY}(x,y) &= f_{Y \mid X = x}(y) \cdot f_X(x) \\
    &= \frac{1}{\sigma_N} \varphi \left(\frac{x - y}{\sigma_N}\right) \cdot \exp(-\theta(x)) \\
    &\propto \exp\left(-\frac{(x - y)^2}{2\sigma_N^2} - \theta(x)\right), \quad x \in \bR,
\end{align*}
which yields the posterior density
\begin{align}
\begin{split}
\label{eq:posterior_strong_log_concave}
    f_{X \mid Y = y}(x) &= \frac{f_{XY}(x,y)}{f_Y(y)} \\
    &\propto \exp\left(-\frac{(x - y)^2}{2\sigma_N^2} - \theta(x)\right), \quad x \in \bR.
\end{split}
\end{align}
Here, we treat $f_{X \mid Y = y}(x)$ as a function of $x$ for fixed $y$, so the marginal $f_Y(y)$ acts as a constant. From~\eqref{eq:posterior_strong_log_concave}, it is clear that $f_{X \mid Y = y}$ is $\sigma_N$-strongly log-concave. However, this characterization alone does not yield a sufficiently strong bound on the posterior variance for Theorem~\ref{thm:Gauss_mech_envelope}. In particular, setting $G(x) = x$ in~\eqref{eq:variance_bound_strong_log_concave} gives
\begin{equation*}
    \operatorname{Var}[X \mid Y = y] \leq \sigma_N^2,
\end{equation*}
whereas our goal is to ensure that $\operatorname{Var}[X \mid Y = y] \leq 0.75 \sigma_N^2$. To close this gap, we must impose a stronger condition on $f_X$. We formalize this requirement in the following corollary.

\begin{cor}
\label{cor:brascamp_var_bound}
Consider the setup of Theorem~\ref{thm:Gauss_mech_envelope}. Let $f_X = \exp(-\theta)$ be the density of $X$, where $\theta$ is twice continuously differentiable and $f_X$ is $\beta$-strongly log-concave. If $\beta \leq \sqrt{3 \sigma_N^2}$, then $\varepsilon_d (\delta) = \log 
\frac{2}{\delta}$, for sufficiently small $\delta \in (0,1)$. 
\end{cor}
\begin{IEEEproof}
We only need to confirm that $\beta \leq \sqrt{3 \sigma_N^2}$ ensures $\operatorname{Var}[X \mid Y=y] \leq 0.75 \sigma_N^2$ for all $y \in \bR$. Let   $f_{X \mid Y=y}(x) \propto \exp (-\gamma(x))$ with 
\begin{gather*}
    \gamma(x) = \frac{(x-y)^2}{2 \sigma_N^2} + \theta(x),\\
    \gamma''(x) = \frac{1}{\sigma_N^2} + \theta''(x) \geq \frac{1}{\sigma_N^2} + \frac{1}{\beta^2} \geq \frac{4}{3 \sigma_N^2}.
\end{gather*}
Then, applying \eqref{eq:variance_bound_strong_log_concave} yields $\operatorname{Var}[X \mid Y=y] \leq 0.75 \sigma_N^2$ for all $y \in \bR$, as desired. 
\end{IEEEproof}

\section{Discussion and Conclusions}
We conclude the paper with some remarks. First, observe that using Corollary~\ref{cor:brascamp_var_bound}, we can view Theorem~\ref{thm:Gauss_mech_Gauss_secret_envelope} as a special case of Theorem~\ref{thm:Gauss_mech_envelope}: If $X \sim \cN (0, \sigma_X^2)$, then $f_X$ is $\sigma_X$-strongly log concave, and the condition required by Corollary~\ref{cor:brascamp_var_bound} becomes $\sigma_X^2 \leq 3\sigma_N^2$. 

Second, the conditions of both Theorem~\ref{thm:Gauss_mech_Gauss_secret_envelope} and Corollary~\ref{cor:brascamp_var_bound} can be interpreted as a concentration requirement on the prior. In particular, when $X$ is $\beta$-strongly log-concave, a smaller $\beta$ corresponds to a more tightly concentrated prior. This requirement is reminiscent of the Gaussian mechanism under differential privacy, where a smaller global sensitivity leads to stronger privacy guarantees.

Future work includes characterizing when $\varepsilon_c(\delta)=\varepsilon_d(\delta)$ and when a gap exists between these two quantities. Other directions involve studying the envelope outside the regimes covered by
Theorems~\ref{thm:Gauss_mech_Gauss_secret_envelope} and \ref{thm:Gauss_mech_envelope}, as well as deriving the envelope when $X$ has bounded support.

\clearpage
\printbibliography

\clearpage
\onecolumn
\appendices
\section{Properties of the Gaussian Mechanism}
\label{ssec:lemmata}
In \cite[Cor.~2]{dytsoConditionalMeanEstimation2023}, it was shown that the information density of the Gaussian mechanism is concave. For completeness, and because several steps of the argument are later reused, we include a proof here.
\begin{lemma}
\label{lemma:info_density_concave}
For the Gaussian mechanism and for each $x \in \supp(P_X)$, the mapping $y \mapsto i(x;y)$ is strictly concave.     
\end{lemma}
\begin{IEEEproof}
Fix $x \in \supp(P_X)$. The information density is 
\begin{equation*}
    i(x;y) = \log \frac{f_{Y \mid X=x}(y)}{f_Y(y)} = \frac{\frac{1}{\sigma_N} \, \varphi \big(\frac{y-x}{\sigma_N} \big)}{f_Y(y)}. 
\end{equation*}
Since $i(x;y)$ is infinitely differentiable, we show that the second derivative is negative. We write 
\begin{subequations}
\begin{align}
    \frac{\partial}{\partial y} i(x;y) &= \frac{\partial}{\partial y} \Big(\log \; \frac{f_{Y \mid X=x}(y)}{f_Y(y)} \Big) \nonumber\\[0.5em]
    &= \frac{\frac{\partial}{\partial y} f_{Y \mid X=x}(y)}{f_{Y \mid X=x}(y)} - \frac{f'_Y(y)}{f_Y(y)} \nonumber\\[0.5em]
    &= \frac{\frac{\partial}{\partial y} \varphi \big(\frac{y-x}{\sigma_N} \big)}{\varphi \big(\frac{y-x}{\sigma_N} \big)} -  \frac{f'_Y(y)}{f_Y(y)} \nonumber \\[0.5em] 
    &= \frac{x-y}{\sigma_N^2} - \frac{\bE[X \mid Y=y] -y}{\sigma_N^2} \label{subeq:tweedie} \\[0.5em]
    & = \frac{x - \bE[X \mid Y=y]}{\sigma_N^2}, \nonumber
\end{align}
\end{subequations}
where $\eqref{subeq:tweedie}$ is due to \emph{Tweedie's formula} for the Gaussian noise channel~\cite{robbins1992empirical, efron2011tweedie} stated as  
\begin{equation}
\label{eq:tweedie}
    \bE[X \mid Y=y] = \sigma_N^2 \, \frac{f'_Y(y)}{f_Y(y)} + y, \quad Y \mid X=x \sim \cN(x, \sigma_N^2). 
\end{equation}
Next, we calculate the second derivative 
\begin{align}
\begin{split}
\label{eq:derivative_posterior_mean}
    \frac{\partial^2}{\partial^2 y} i(x;y) &= -\frac{1}{\sigma_N^2} \cdot \frac{\partial}{\partial y} \bE[X \mid Y=y]\\
    &= -\frac{1}{\sigma_N^2} \cdot \frac{\partial}{\partial y} \int x \, f_{X \mid Y=y}(x) \, dx\\[0.5em]
    &= -\frac{1}{\sigma_N^2} \cdot \frac{\partial}{\partial y} \int x \, f_X(x) \frac{f_{Y \mid X=x}(y)}{f_Y(y)}\, dx\\[0.5em]
    &= -\frac{1}{\sigma_N^2} \int x \, f_X(x) \left(\frac{f_{Y \mid X=x}(y)}{f_Y(y)} \right) \left( \frac{\frac{\partial}{\partial y} f_{Y \mid X=x}(y)}{f_{Y \mid X=x}(y)} - \frac{f'_Y(y)}{f_Y(y)} \right) \, dx\\[0.5em] 
    &= -\frac{1}{\sigma_N^4} \int x \, \left(x - \bE[X \mid Y=y] \right) \, f_{X \mid Y=y}(x) \, dx\\[0.5em]
    &= - \frac{\mathrm{Var} [X \mid Y=y]}{\sigma_N^4} <0. 
\end{split}    
\end{align}
Therefore, $y \mapsto i(x;y)$ is concave for all $x$. 
\end{IEEEproof}
An important consequence of Lemma~\ref{lemma:info_density_concave} is that for fixed $x$, the super-level sets of $i(x;y)$ are convex.

\subsection{Proof of Lemma~\ref{lemma:tail_worse}}
\label{ssec:lemma_tail_worse_proof}
\begin{enumerate}
\item By \cite[Lemma 1(i)]{csf_envelope}, $\log \frac{1}{\delta}$ is the largest amount of information leakage for any set with probability at least $\delta$. Therefore, it suffices to verify that $(-\infty, t_L)$ and $(t_R, \infty)$ attain this bound. We have 
\begin{align*}
    \ell(X \to (-\infty, t_L)) &= \log \; \sup_{x \in \bR} \; \frac{P_{Y \mid X=x} (-\infty, t_L)}{P_{Y} (-\infty, t_L)}\\
    &= \log \; \sup_{x \in \bR} \; \frac{\Phi(\frac{t_L - x}{\sigma_N})}{\delta}\\
    &= \log \frac{1}{\delta}, 
\end{align*}
since $\lim_{x \to -\infty} \Phi(\frac{t_L - x}{\sigma_N}) = 1$. Similarly, it can be shown that $\ell(X \to (t_R, \infty)) = \log \frac{1}{\delta}$. 

\item 
By Lemma \cite[Lemma 1(iv)]{csf_envelope}, we can restrict the optimization over $\cE$ to sets with probability exactly $\delta$. As such, we have 
\begin{align}
\begin{split}    
\label{eq:main_opt_prob}
    &\sup_{\cE \subset (a,b) : P_Y(\cE) \geq \delta} \; \ell (X \to \cE) =\\
    &\hspace{4em}\log \; \sup_{\cE \subset (a,b) : P_Y(\cE) = \delta} \; \sup_{x \in \bR}  \; \frac{P_{Y \mid X=x} (\cE)}{\delta}. 
\end{split}    
\end{align}    
Below, we lower and upper bound 
\begin{equation*}
    \sup_{\cE \subset (a,b) : P_Y(\cE) = \delta} \; \sup_{x \in \bR} \;  P_{Y \mid X=x} (\cE).
\end{equation*}

\underline{Lower bound:}
Let $\cI_{(a,b)}$ denote the collection of all intervals in $(a,b)$, and observe that 
\begin{equation}
\label{eq:intervals_lame} 
    \sup_{\cE \in \cI_{(a,b)} : P_Y(\cE) = \delta} \; \sup_{x \in \bR} \; P_{Y \mid X=x}(\cE) \leq\sup_{\cE \subset (a,b) : P_Y(\cE) = \delta} \; \sup_{x \in \bR} \; P_{Y \mid X=x}(\cE),
\end{equation}
since $\cI_{(a,b)}$ is a sub-collection of all the measurable sets in $(a,b)$. Fix an interval $\cA = (c,d) \in \cI_{(a,b)}$, and observe that
\begin{equation}
\label{eq:interval_cond_prob_Gaussian}
    \sup_{x \in \bR} \, P_{Y \mid X=x}(\cA) = \sup_{x \in \bR} \; \Big(\Phi(\frac{d-x}{\sigma_N}) - \Phi(\frac{c-x}{\sigma_N}) \Big).
\end{equation}
The mapping $x \mapsto \Phi(\frac{b-x}{\sigma_N}) - \Phi(\frac{a-x}{\sigma_N})$ while not concave, is unimodal and differentiable. Thus, we can find its maximizer $x^*$ by setting its derivative equal to zero:
\begin{equation}
\label{eq:best_x_middle}
    \varphi(\frac{d-x^*}{\sigma_N}) = \varphi(\frac{c-x^*}{\sigma_N}) \implies x^* = \frac{c+d}{2}. 
\end{equation}
Plugging $x^*$ back into~\eqref{eq:interval_cond_prob_Gaussian} yields 
\begin{equation*}
     \sup_{x \in \bR} \, P_{Y \mid X=x}(\cA) = 2 \Phi(\frac{d-c}{2 \sigma_N}) - 1 = 2 \Phi(\frac{\abs{\cA}}{2 \sigma_N}) - 1, 
\end{equation*}
where $\abs{\cA}$ denotes the Lebesgue measure of $\cA$. Combining the above equality with \eqref{eq:intervals_lame}, we get 
\begin{equation}
\label{eq:lower_bound_intervals}
    \sup_{\cE \subset (a,b) : P_Y(\cE) = \delta} \; \sup_{x \in \bR} \; P_{Y \mid X=x}(\cE) \geq\sup_{\cA \in \cI_{(a,b)} : P_Y(\cA) = \delta} \; 2 \Phi(\frac{\abs{\cA}}{2 \sigma_N}) - 1.   
\end{equation}

\underline{Upper bound:} Fix $x \in \bR$, and consider the optimization problem: 
\begin{align}
\begin{split}
\label{eq:conditonal_dist_optimization}
    \sup_{\cE \subset (a,b)} \; &P_{Y \mid X=x}(\cE),\\
    \text{s.t.} \; &P_Y(\cE) = \delta. 
\end{split}
\end{align}
By the bathtub principle~\cite[Thm. 14]{lieb2001analysis}, or alternatively, the Neyman-Pearson lemma~\cite{neyman1933ix}, the optimal set $\cE^*_x$ is a super-level set of the information density, that is, 
\begin{equation*}
    \cE^*_x = \big\{y \in (a,b) : i(x;y) > \tau^* \big\},
\end{equation*}
where $\tau^*$ is selected such that 
\begin{equation*}
    P_Y(\cE^*_x) = \delta. 
\end{equation*}
(See Appendix~\ref{ssec:proof_bathtub_1} for the proof of the optimality of $\cE^*_x$.) Furthermore, due to the concavity of $y \mapsto i(x;y)$ for fixed $x$ (Lemma~\ref{lemma:info_density_concave}), the set $\cE^*_x$ is an interval since super-level sets of concave functions are convex. 

Now, suppose $\cE^*_x = (e_x, f_x) \subset (a,b)$. We write 
\begin{subequations}
\begin{align}
    \sup_{x \in \bR} \; \sup_{\cE \subset (a,b) : P_Y(\cE) = \delta} \; P_{Y \mid X=x}(\cE) &= \sup_{x \in \bR} \; P_{Y \mid X=x}(\cE_x^*)\nonumber\\
    &=  \sup_{x \in \bR} \; \Big(\Phi(\frac{f_x-x}{\sigma_N}) - \Phi(\frac{e_x-x}{\sigma_N}) \Big)\nonumber\\
    &\leq \sup_{x \in \bR}\; \sup_{z \in \bR} \; \Big( \Phi(\frac{f_x-z}{\sigma_N}) - \Phi(\frac{e_x-z}{\sigma_N}) \Big)\nonumber\\
    &= \sup_{x \in \bR} \; 2 \Phi(\frac{f_x-e_x}{2\sigma_N})- 1 \label{subeq:opt_interval}\\
    &\leq \sup_{\cA \in \cI_{(a,b)} : P_Y(\cA) = \delta} \; 2 \Phi(\frac{\abs{\cA}}{2 \sigma_N}) - 1, \label{subeq:upper_bound_intervals}
\end{align}
\end{subequations}
where~\eqref{subeq:opt_interval} follows by a calculation similar to \eqref{eq:best_x_middle}. Then, putting together~\eqref{subeq:upper_bound_intervals} and \eqref{eq:lower_bound_intervals}, we conclude that 
\begin{align}
\begin{split}
\label{eq:equivalent_optimization}
    \sup_{x \in \bR} \; \sup_{\cE \subset (a,b) : P_Y(\cE)= \delta} \; P_{Y \mid X=x}(\cE) = \sup_{\cA \in \cI_{(a,b)} : P_Y(\cA) = \delta} \; 2 \Phi(\frac{\abs{\cA}}{2 \sigma_N}) - 1. 
\end{split}
\end{align}

Next, observe that the RHS of~\eqref{eq:equivalent_optimization} is increasing in $\abs{\cA}$. Thus, to solve~\eqref{eq:equivalent_optimization}, we can alternatively solve
\begin{align}
\begin{split}
\label{eq:lebesgue_optimization}
     \sup_{\cA \in \cI_{(a,b)}} \; \abs{\cA},\\
    \text{s.t.} \; P_Y(\cA) = \delta. 
\end{split}    
\end{align}
Let
\begin{gather*}
    \cK_\delta \;=\; \bigl\{(u,v)\in [a,b]^2 : F_Y(v)-F_Y(u)=\delta \bigr\},\\
    L(u,v)=v-u ,
\end{gather*}
where $F_Y$ is the CDF of $Y$, and note that~\eqref{eq:lebesgue_optimization} can be expressed as 
\begin{equation*}
    \sup_{(u,v)\in\mathcal \cK_\delta} \; L(u,v).
\end{equation*}
The map $(u,v)\mapsto F_Y(v)-F_Y(u)$ is continuous, and since the singleton $\{\delta\}$ is a closed subset of $\bR$, its pre-image under $F_Y(v)-F_Y(u)$ is a closed (and non-empty) subset of $\cK_\delta$. As $\cK_\delta$ is also bounded, we conclude that it is a compact set. Moreover, the length function $L$ is continuous on $\cK_\delta$. Hence, by the extreme-value theorem, the suprema in both \eqref{eq:lebesgue_optimization} and \eqref{eq:equivalent_optimization} are attained. It follows that there exists an interval $\cA^* \in \cI_{(a,b)}$ such that 
\begin{equation*}
    \ell (X \to \cA^*) = \sup_{\cE \subset (a,b) : P_Y(\cE) \geq \delta} \; \ell (X \to \cE), 
\end{equation*}
as desired.

\item 
Given $a,b \in \bR$, let $a_\delta, b_\delta \in (a,b)$ satisfy
\[
P_Y(a,b_\delta) = P_Y (a_\delta,b) = \delta,
\]
and define the corresponding density thresholds $\tau_L := f_Y(b_\delta)$ and $\tau_R := f_Y(a_\delta)$. Suppose $a$ is sufficiently smaller than $-M$, $b$ is sufficiently larger than $M$, and $\delta$ is sufficiently small so that $b_\delta < -M$ and $a_\delta > M$. 

We now introduce an auxiliary ``core'' interval $(a_1,b_1)\subset(a,b)$ whose purpose is to separate the left and right sub-level sets of $f_Y$. Since $f_Y$ is strictly increasing on $(-\infty,-M)$ and strictly decreasing on $(M,\infty)$ (and in particular $f_Y(y)\to 0$ as $|y|\to\infty$), for $a$ sufficiently smaller than $-M$, $b$ sufficiently larger than $M$, and $\delta$ sufficiently small, there exists an interval $(a_1, b_1) \subset (a,b)$ satisfying
\begin{gather}
    P_Y (a_1, b_1) > \delta, \nonumber\\ 
    \big\{y \in (a, b_1) : f_Y(y) < \tau_L \big\} = (a,b_\delta),\label{eq:a_1_b_1}\\ 
    \big\{y \in (a_1, b) : f_Y(y) < \tau_R \big\} = (a_\delta,b).\nonumber  
\end{gather}
See Figure~\ref{fig:sub_level_tail} for an illustration of this construction.

Observe that all intervals with probability $\delta$ are either entirely contained in $(a,b_1)$ or in $(a_1,b)$.\footnote{Let $(u,v) \subset (a,b)$. If $u \leq a_1$ and $v \geq b_1$, then $(a_1,b_1)\subset(u,v)$ and hence $P_Y(u,v) \geq P_Y (a_1,b_1) > \delta$.} Thus, we can find the solution to \eqref{eq:lebesgue_optimization} by solving the problem over $\cI_{(a,b_1)}$ and $ \cI_{(a_1,b)}$ separately and then taking the larger value, i.e.,
\begin{align}
\begin{split}
\label{eq:two_prob_max}
    \sup_{\substack{\cA \in \cI_{(a,b)} :\\ P_Y(\cA) = \delta}} \; \abs{\cA}
    =\max \Big\{\sup_{\substack{\cA \in \cI_{(a,b_1)} :\\ P_Y(\cA) = \delta}} \; \abs{\cA}, \;
    \sup_{\substack{\cA \in \cI_{(a_1,b)} :\\ P_Y(\cA) = \delta}} \; \abs{\cA} \Big\}.  
\end{split}    
\end{align}

Now, we solve the problem over $(a,b_1)$. Consider the following relaxation:
\begin{align}
\begin{split}
\label{eq:relaxed_optimization_2}
     \sup_{\cA \subset (a,b_1)} \; \abs{\cA},\\
    \text{s.t.} \; P_Y(\cA) = \delta, 
\end{split}
\end{align}
where we have enlarged the feasible set from all the intervals in $(a,b_1)$ to all measurable subsets of $(a,b_1)$. By the bathtub principle, the solution to \eqref{eq:relaxed_optimization_2} is a sub-level set of the density function $f_Y$, i.e.,
\begin{equation*}
    \cA^*_{(a,b_1)} = \Big\{y \in (a, b_1) : f_Y(y) < \tau^* \Big\}, 
\end{equation*}
where $\tau^*$ is picked such that $P_Y(\cA^*_{(a,b_1)}) = \delta$. (See Appendix~\ref{ssec:proof_bathtub_2} for the proof of the optimality of $\cA^*_{(a,b_1)}$.) However, by \eqref{eq:a_1_b_1}, the unique sub-level set inside $(a,b_1)$ achieving probability $\delta$ is the left-tail interval adjacent to $a$, namely
\begin{gather*}
      \cA^*_{(a,b_1)} = \Big\{y \in (a, b_1) : f_Y(y) < \tau_L \Big\} = (a, b_\delta),\\
      P_Y (a, b_\delta) = \delta.
\end{gather*}
Since the solution to \eqref{eq:relaxed_optimization_2} is an interval, it is also the solution to the first problem in the RHS of \eqref{eq:two_prob_max}.

Applying the same argument to $(a_1,b)$, we get that the solution to the second problem in the RHS of \eqref{eq:two_prob_max} is the right tail interval adjacent to $b$, that is,
\begin{gather*}
      \cA^*_{(a_1,b)} = \Big\{y \in (a_1, b) : f_Y(y) < \tau_R \Big\} = (a_\delta, b),\\
      P_Y (a_\delta, b) = \delta. 
\end{gather*}
We conclude that either $(a_\delta, b)$ or $(a, b_\delta)$ is the solution to the original problem \eqref{eq:lebesgue_optimization}.

\begin{figure}[t]
    \centering
    \includegraphics[scale=0.8]{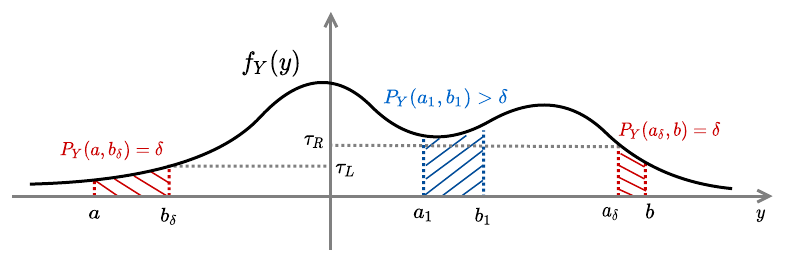}
    \caption{An example illustrating the points $a_1, b_1, a_\delta$ and $b_\delta$.}
    \label{fig:sub_level_tail}
\end{figure}
\end{enumerate}

\section{Applications of the Bathtub Principle} 
\begin{theorem}[{Bathtub principle~\cite[Thm. 1.14]{lieb2001analysis}}]
\label{thm:bathtub}
Let $(\Omega, \Sigma, \mu)$ be a $\sigma$-finite measure space and let $f : \Omega \to \mathbb{R}$ be a real-valued measurable function such that $\mu(\{\omega : f(\omega) < t\}) < \infty$ for all $t \in \mathbb{R}$. Let $G > 0$ and define the class of measurable functions:
\begin{equation*}
    \mathcal{C} = \left\{ g : \Omega \to [0,1] \ \middle| \ \int_\Omega g(\omega)\,\mu(d\omega) = G \right\}.
\end{equation*}
Then the minimization problem
\begin{equation*}
\inf_{g \in \mathcal{C}} \int_\Omega f(\omega) g(\omega)\,\mu(d\omega)
\end{equation*}
is solved by
\begin{equation*}
g(\omega) = \ind_{\{f < s\}}(\omega) + c \, \ind_{\{f = s\}}(\omega),
\end{equation*}
where
\begin{equation*}
s = \sup \left\{ t \in \mathbb{R} : \mu(\{f < t\}) \leq G \right\}
\quad \text{and} \quad
c \, \mu(\{f = s\}) + \mu(\{f < s\}) = G.
\end{equation*}
The minimizer $g$ is unique if $G = \mu(\{f < s\})$ or $G = \mu(\{f \leq s\})$.
\end{theorem}

\subsection{Finding the Solution of \texorpdfstring{\eqref{eq:conditonal_dist_optimization}}{Eq.~\ref{eq:conditonal_dist_optimization}}}
\label{ssec:proof_bathtub_1}
Let $(a,b) \subset \bR$ and $\delta \in (0,1)$. Fix $x \in \bR$, and consider the problem 
\begin{align*}
    \sup_{\cE \subset (a,b)} \; P_{Y \mid X=x}(\cE),\\
    \text{s.t.} \; P_Y(\cE) = \delta. 
\end{align*}
To use the bathtub principle, we solve the equivalent problem
\begin{gather*}
    \inf_{\cE \subset (a,b)} \left( - \int_\cE f_{Y \mid X=x}(y) \, dy\right) =   \inf_{\cE \subset (a,b)} \, \int_{(a,b)} \ind_\cE(y) \left( - \frac{f_{Y \mid X=x}(y)}{f_Y(y)}\right) \,P_Y(dy), \\[0.6em]
    \text{s.t.} \; P_Y(\cE) = \int_{(a,b)} \, \ind_\cE(y) \, P_Y(dy) = \delta. 
\end{gather*}
By Theorem~\ref{thm:bathtub}, the solution $\cE^*_x$ is 
\begin{equation*}
    \cE^*_x = \big\{y \in (a,b) : - \frac{f_{Y \mid X=x}(y)}{f_Y(y)} < \tau^* \big\} = \big\{y \in (a,b) : i(x;y) > \log(-\tau^*) \big\},
\end{equation*}
where $\tau^*$ is unique and satisfies $P_Y(\cE^*_x) = \delta$. Note that since $i(x;y)$ is real analytic, all of its level sets have probability zero. 

\subsection{Finding the Solution of \texorpdfstring{\eqref{eq:relaxed_optimization_2}}{Eq.~\ref{eq:relaxed_optimization_2}}}
\label{ssec:proof_bathtub_2}
Let $(a,b) \subset \bR$, $\delta \in (0,1)$ and consider the problem
\begin{align*}
     \sup_{\cA \subset (a,b)} \; \abs{\cA},\\
    \text{s.t.} \; P_Y(\cA) = \delta. 
\end{align*}
To find the optimal set, we may solve the equivalent problem 
\begin{gather*}
     \inf_{\cA \subset (a,b)} \; \left(- \int_{(a,b)} \ind_{\cA}(y) \, dy \right) = \inf_{\cA \subset (a,b)} \;  \int_{(a,b)} \ind_{\cA}(y) \, \left(- \frac{1}{f_Y(y)} \right) \, P_Y(dy)  ,\\
    \text{s.t.} \; P_Y(\cA) =  \int_{(a,b)} \ind_{\cA}(y) \, P_Y(dy) = \delta. 
\end{gather*}
By Theorem~\ref{thm:bathtub}, the solution is $\cA^*$ expressed by 
\begin{equation*}
    \cA^* = \Big\{y \in (a, b) : - \frac{1}{f_Y(y)} < \tau^* \Big\} = \Big\{y \in (a, b) : f_Y(y) < -\frac{1}{\tau^*} \Big\}, 
\end{equation*}
where $\tau^*$ is unique and satisfies $P_Y(\cE^*_x) = \delta$.

\section{Proofs for the Gaussian Mechanism with a Gaussian Secret}

\subsection{Proof of Lemma~\ref{lemma:gaussian_interval_leakage}}
We have 
\label{ssec:gaussian_interval_leakage_proof}
\begin{align}
\begin{split}
\label{eq:event_pml_Gaussian_2}
    \ell(X \to (a,b)) &= \log \; \frac{\sup_{x \in \bR} P_{Y \mid X=x} (a,b) }{P_Y(a,b)}\\
    &= \log \; \frac{\sup_{x \in \bR} \Phi(\frac{b-x}{\sigma_N}) - \Phi(\frac{a-x}{\sigma_N})}{\Phi(\frac{b}{\sigma_Y}) - \Phi(\frac{a}{\sigma_Y})}. 
\end{split}    
\end{align}
The mapping $x \mapsto \Phi(\frac{b-x}{\sigma_N}) - \Phi(\frac{a-x}{\sigma_N})$ is unimodal and differentiable. So, we find its maximizer $x^*$ by setting the derivative equal to zero:
\begin{equation*}
    \varphi(\frac{d-x^*}{\sigma_N}) = \varphi(\frac{c-x^*}{\sigma_N}) \implies x^* = \frac{c+d}{2}. 
\end{equation*}
Plugging $x^*$ back into~\eqref{eq:event_pml_Gaussian_2} yields \eqref{eq:event_pml_Gaussian_1}. 

\subsection{Proof of Lemma~\ref{lemma:interval_leakage_increasing}}
\label{ssec:proof_lemma_interval_leakage_increasing}
Let $0 < a < b$. First, we calculate the derivative of $\ell(X \to (a,b))$ in $b$, denoted by $\ell'(X \to (a,b))$:
\begin{align}
\label{eq:event_leakage_Guassian_derivative}
    \ell'(X \to (a,b)) &=  \frac{d}{db} \left(\log \, \Big( 2\Phi(\frac{b-a}{2 \sigma_N}) -1 \Big) - \log \Big(\Phi(\frac{b}{\sigma_Y}) - \Phi(\frac{a}{\sigma_Y})\Big) \right)\nonumber\\
    &= \frac{\varphi(\frac{b-a}{2\sigma_N})}{\sigma_N \left(2\Phi(\frac{b-a}{2 \sigma_N}) -1\right)} - \frac{\varphi(\frac{b}{\sigma_Y})}{\sigma_Y \left(\Phi(\frac{b}{\sigma_Y}) - \Phi(\frac{a}{\sigma_Y})\right)}. 
\end{align}

The following lemma, proved in Appendix~\ref{ssec:proof_lemma_event_leakage_first_increase}, shows that $\ell(X \to (a,b))$ at least initially increases in $b$, that is $\ell'(X \to (a,b)) > 0$ when $b$ is sufficiently close to $a$. 

\begin{lemma} 
\label{lemma:event_leakage_first_increase}
Let $a > 0$. There exists $\xi_0 > 0$ such that the mapping $b \mapsto \ell(X \to (a, b))$ is strictly increasing on the interval $(a, a + \xi_0)$. 
\end{lemma}

Let $\sigma_Y^2 = \sigma_X^2 +  \sigma_N^2 $ denote the variance of $Y$. Next, we argue that when $3 \sigma_N^2 \geq \sigma_X^2$, equivalently $4\sigma_N^2 \geq \sigma_Y^2$, then $\ell'(X \to (a,b)) > 0$ for all $b > a$. To prove this, let  
\begin{equation*}
    r(b) \coloneqq \frac{\frac{1}{\sigma_N} \varphi(\frac{b-a}{2\sigma_N})}{2\Phi(\frac{b-a}{2\sigma_N}) -1}, 
\end{equation*}
and observe that its derivative is 
\begin{align}
    r'(b) &= - \frac{\frac{(b-a)}{4\sigma_N^3} \varphi(\frac{b-a}{2\sigma_N}) \Big(2 \Phi(\frac{b-a}{2\sigma_N}) -1 \Big) + \frac{1}{\sigma_N^2} \varphi^2(\frac{b-a}{2\sigma_N})}{\left(2\Phi(\frac{b-a}{2\sigma_N}) -1\right)^2}\nonumber\\[0.6em]
    &= - \frac{(b-a)}{4\sigma_N^2} r(b) - r^2(b),\label{eq:r_prime} 
\end{align}
where we used $\frac{d}{dt} \varphi(t) = -t \varphi(t)$ for all $t \in \bR$. Similarly, let 
\begin{equation*}
    s(b) \coloneqq \frac{\frac{1}{\sigma_Y} \varphi(\frac{b}{\sigma_Y})}{\Phi(\frac{b}{\sigma_Y}) - \Phi(\frac{a}{\sigma_Y})},
\end{equation*}
and observe that we can express its derivative by 
\begin{align}
    s'(b) &= \frac{- \varphi(\frac{b}{\sigma_Y})}{\sigma_Y^2 \big(\Phi(\frac{b}{\sigma_Y}) - \Phi(\frac{a}{\sigma_Y})\big)^2} \Big[\frac{b}{\sigma_Y} \big(\Phi(\frac{b}{\sigma_Y}) - \Phi(\frac{a}{\sigma_Y})\big) +  \varphi(\frac{b}{\sigma_Y}) \Big]\nonumber\\
    &= - \frac{b}{\sigma_Y^2} s(b) - s^2(b),\label{eq:s_prime} 
\end{align}
For simplicity, let $u(b) \coloneqq \ell'(X \to (a,b)) = r(b) - s(b)$ which means $u'(b) = r'(b) - s'(b)$. Plugging in \eqref{eq:r_prime} and \eqref{eq:s_prime} into the expression for $u'(b)$ yields 
\begin{align*}
    u'(b) &= - \frac{(b-a)}{4\sigma_N^2} r(b) - r^2(b) + \frac{b}{\sigma_Y^2} s(b) + s^2(b)\\
    &= - \frac{(b-a)}{4\sigma_N^2} \big(u(b) + s(b)\big) - \big(u(b)+s(b)\big)^2 + \frac{b}{\sigma_Y^2} s(b) + s^2(b)\\
    &= -u^2(b) - \Big(2s(b) + \frac{b-a}{4\sigma_N^2} \Big) u(b) + \Big(\frac{b}{\sigma_Y^2}-\frac{b-a}{4\sigma_N^2}\Big) s(b).  
\end{align*}
Now, suppose there exists $b_0 \in (a,\infty)$ with $u(b_0) = \ell'(X \to (a,b_0)) = 0$. At this point, we would have 
\begin{equation*}
    u'(b_0) = \Big(\frac{b_0}{\sigma_Y^2}-\frac{b_0-a}{4\sigma_N^2}\Big) s(b_0). 
\end{equation*}
Note that if $4\sigma_N^2 \geq \sigma_Y^2$, then we have
\begin{equation}
\label{eq:cross_positive}
    \frac{b}{\sigma_Y^2}-\frac{b-a}{4\sigma_N^2} \geq \frac{b}{\sigma_Y^2}-\frac{b-a}{\sigma_Y^2} =  \frac{a}{\sigma_Y^2} >0, \quad b \in (a,\infty),
\end{equation}
and also $s(b)>0$ for all $b \in (a,\infty)$. Therefore, $u'(b_0) >0$, that is, the function $u(b)$ would always cross zero with a positive slope, changing sign from negative to positive. By the continuity of $u(b)$, this means that there exists $b_1 < b_0$ with $u(b_1) <0$. On the other hand, Lemma~\ref{lemma:event_leakage_first_increase} states that $u(b)$ is positive for $b$ sufficiently close to $a$, which in turn, implies that $u(b)$ must have crossed zero and changed sign from positive to negative at some point before $b_0$. That is, there exists  $b_2 < b_0$ such that $u(b_2) =0$, $u(b_2^-) >0$ and  $u(b_2^+) <0$. However, by \eqref{eq:cross_positive} this is impossible since $u(b)$ can only cross zero with a positive slope. 

We conclude that if $4\sigma_N^2 \geq \sigma_Y^2$, then $u(b) = \ell'(X \to (a,b)) >0$ for all $b \in (a,\infty)$. Thus, the leakage $\ell(X \to (a,b))$ is increasing and approaches its limits which is 
\begin{equation*}
    \lim_{b \to \infty} \ell(X \to (a,b)) = \log \frac{1}{P_Y(a,\infty)}. 
\end{equation*}

\subsection{Proof of Lemma~\ref{lemma:event_leakage_first_increase}}
\label{ssec:proof_lemma_event_leakage_first_increase}
Fix $a > 0$. We show that the derivative in \eqref{eq:event_leakage_Guassian_derivative} is positive at $b = a + \xi$, or equivalently, 
\begin{equation*}
   \sigma_Y \, \varphi(\frac{\xi}{2\sigma_N}) \cdot \left(\Phi(\frac{a + \xi}{\sigma_Y}) - \Phi(\frac{a}{\sigma_Y})\right) - \sigma_N \, \varphi(\frac{a + \xi}{\sigma_Y}) \cdot \left(2\Phi(\frac{\xi}{2 \sigma_N}) -1\right) > 0,
\end{equation*}
for sufficiently small $\xi$. For this, we use the Taylor expansions 
\begin{gather*}
     \varphi(\frac{\xi}{2\sigma_N}) = \varphi(0) \left(1 - \frac{1}{8 \sigma_N^2} \xi^2 \right) + O(\xi^4), \\
     \Phi(\frac{a + \xi}{\sigma_Y}) -  \Phi(\frac{a}{\sigma_Y}) =  \frac{\varphi(\frac{a}{\sigma_Y})}{\sigma_Y} \xi - \frac{a \varphi(\frac{a}{\sigma_Y})}{2 \sigma_Y^3} \xi^2 + O(\xi^3),\\
     \varphi(\frac{a + \xi}{\sigma_Y}) = \varphi(\frac{a}{\sigma_Y}) \Bigg(1 - \frac{a}{\sigma_Y^2} \xi + \frac{\big(\frac{a}{\sigma_Y}\big)^2 -1}{2 \sigma_Y^2} \xi^2 \Bigg) + O(\xi^3), \\
     2\Phi(\frac{\xi}{2 \sigma_N}) -1 = \frac{\varphi(0)}{\sigma_N}\xi-
        \frac{\varphi(0)}{24\,\sigma_N^{3}}\,\xi^{3}
        + O(\xi^{5}).
\end{gather*}
We have 
\begin{equation*}
    \sigma_Y \, \varphi(\frac{\xi}{2\sigma_N}) \cdot \left(\Phi(\frac{a + \xi}{\sigma_Y}) - \Phi(\frac{a}{\sigma_Y})\right) - \sigma_N \, \varphi(\frac{a + \xi}{\sigma_Y}) \cdot \left(2\Phi(\frac{\xi}{2 \sigma_N}) -1\right)  = \varphi(0)\,\varphi(\frac{a}{\sigma_Y}) \big(\frac{a}{2\sigma_Y^{2}} \big)\xi^{2} + O(\xi^{3}),
\end{equation*}
which is positive for $a>0$ and sufficiently small $\xi$.

\subsection{Proof of Lemma~\ref{lemma:simple_to_measurable}}
\label{ssec:lemma_simple_to_measurable_proof}
Let $Z = h(Y)$, where $h: \bR \to \cZ$ is an arbitrary measurable function. Consider a sequence of simple functions $k_n : \cZ \to \cZ_n$ that converge pointwise to the identity map $k (z) = z$. Put $Z_n = k_n (Z)$ and define $\ell_n (X \to z) \coloneqq \ell_{P_{XZ^n}}(X \to k_n(z))$. Note that the sequence $(k_n)$ can be constructed in such a way that $\sigma (Z_1) \subset \sigma(Z_2) \subset \cdots$ and $\lim_{n \to \infty} \sigma(Z_n) = \sigma(Z)$, where $\sigma(Z_n)$ denotes the $\sigma$-algebra on $\Omega$ generated by $Z_n$.  

Let $\psi : \cX \to \bR_+$ be a bounded and continuous function, and put $M_n = \bE [\psi(X) \mid \sigma(Z_n) ]$. It can be easily checked that $M_n$ is a martingale (see, e.g.,~\cite[Prop. V.2.7]{cinlar2006probability}). In particular, Doob's martingale convergence theorem~\cite[Thm. V.4.7]{cinlar2006probability} states that 
\begin{equation*}
    M_n \to M \coloneqq \bE [\psi(X) \mid \sigma(\cup_{n=1}^{\infty} \, \sigma(Z_n))]\ = \bE [\psi(X) \mid \sigma(Z)],  \quad \mathrm{a.s..}
\end{equation*}
Since the above holds for any bounded and continuous function, it implies that $P_{X \mid Z_n}$ converges (almost sure) weakly to $P_{X \mid Z}$. We now exploit the fact that the Rényi divergence is lower semi-continuous in the topology of weak convergence~\cite[Thm. 19]{van2014renyi}, i.e., given a sequence of probability measures $P_n$ that converge weakly to $P$ and a fixed probability measure $Q$ we have 
\begin{equation*}
    \liminf_{n \to \infty} \dinf (P_n \Vert Q) \geq \dinf (P \Vert Q).  
\end{equation*}
Applying the above, we get
\begin{equation*}
    \liminf_{n \to \infty} \, \ell_n(X \to z) = \liminf_{n \to \infty} \, \dinf (P_{X \mid Z_n = k_n(z)} \Vert P_{X}) \geq \dinf (P_{X \mid Z = z}  \Vert P_{X}) = \ell (X \to z), \quad P_Z\mathrm{-a.s.}.  
\end{equation*}
The final step is to consider the implications of the above inequality for the survival functions. For all $t \geq 0$, we have 
\begin{align}
    \bP \{\ell (X \to Z) > t \} &\leq \bP \{\liminf_{n \to \infty} \, \ell_n( X \to Z) > t \} \nonumber\\
    &\leq \liminf_{n \to \infty} \bP \{\ell_n( X \to Z) > t \}, \label{eq:tail_liminf}
\end{align}
where the second inequality follows from Fatou's lemma. Now, take any $t<\bar\varepsilon_Z(\delta)$. By the definition of $\bar\varepsilon_Z(\delta)$,
\begin{equation*}
    \mathbb{P}\bigl\{\ell(X\to Z)>t\bigr\}\ge \delta, 
\end{equation*}
and combining this with \eqref{eq:tail_liminf} yields
\begin{equation*}
    \liminf_{n\to\infty}\mathbb{P}\bigl\{\ell(X\to Z_n)>t\bigr\}\ge \delta.
\end{equation*}
Fix a small $\eta >0$. The above inequality implies that there exists $N=N(t,\eta)$ such that for all $n\ge N$,
\begin{equation*}
    \mathbb{P}\bigl\{\ell(X\to Z_n)>t \bigr\} \geq \delta-\eta,
\end{equation*}
since otherwise $\mathbb{P}\{\ell(X\to Z_n)>t\}\leq \delta-\eta$ for infinitely many $n$, which would imply
$\liminf_{n\to\infty} \mathbb{P}\{\ell(X\to Z_n)>t\}\le \delta-\eta$. By the definition of $\bar\varepsilon_{Z_n}(\delta-\eta)$, this implies that for all $n\ge N$,
\begin{equation*}
    t \leq \bar \varepsilon_{Z_n}(\delta-\eta).
\end{equation*}
Therefore
\begin{equation*}
    t \leq \liminf_{n\to\infty} \bar \varepsilon_{Z_n}(\delta-\eta).
\end{equation*}
Since this holds for every $t<\bar\varepsilon_Z(\delta)$, taking the supremum over such $t$ gives
\begin{equation*}
    \bar\varepsilon_Z(\delta) \leq \liminf_{n\to\infty }\bar \varepsilon_{Z_n}(\delta-\eta).
\end{equation*}
Then, using the bound for simple post-processings, we obtain
\begin{equation*}
    \bar \varepsilon_Z(\delta) \leq \liminf_{n\to\infty}\bar\varepsilon_{Z_n}(\delta-\eta) \leq \log \frac{2}{\delta-\eta},
\end{equation*}
and letting $\eta\downarrow 0$ yields $\bar\varepsilon_Z(\delta)\leq \log\frac{2}{\delta}$.

\section{Proofs for the Gaussian Mechanism with Unbounded Secret}
\subsection{Proof of Theorem~\ref{thm:Gauss_mech_envelope}} 
\label{ssec:thm_Gauss_mech_envelope_proof}
The proof follows the same template as the proof of Theorem~\ref{thm:Gauss_mech_Gauss_secret_envelope}.  Accordingly, we begin by calculating the information leaked to intervals. Let $F_Y$ denote the CDF of $Y$. A derivation similar to that of Lemma~\ref{lemma:gaussian_interval_leakage} yields   
\begin{equation}
    \ell(X \to (a,b)) = \log \; \frac{2\Phi(\frac{b-a}{2 \sigma_N}) -1}{F_Y(b) - F_Y(a)}, \quad (a,b) \subset \bR. 
\end{equation}

The key technical component of the proof of Theorem~\ref{thm:Gauss_mech_Gauss_secret_envelope} was a monotonicity statement for $b \mapsto \ell(X\to(a,b))$ given fixed $a$. The following lemma (proved in Appendix~\ref{ssec:lemma_interval_leakage_increasing_general_proof}) plays the same role as Lemma~\ref{lemma:interval_leakage_increasing}, but is formulated under the more general assumption of an unbounded secret. 

\begin{lemma}
\label{lemma:interval_leakage_increasing_general}   
Suppose $4 \operatorname{Var}[X \mid Y=y] \leq 3 \sigma_N^2$ for all $y \in \bR$. Then for all $a > M$, the mapping $b \mapsto \ell(X \to (a,b))$ is strictly increasing and has the limit 
\begin{equation*}
    \lim_{b \to \infty} \ell(X \to (a,b)) = \log \; \frac{1}{P_Y (a, \infty)}. 
\end{equation*}    
\end{lemma}

Equipped with Lemma~\ref{lemma:interval_leakage_increasing_general}, the proof proceeds along the same lines as that of Theorem~\ref{thm:Gauss_mech_Gauss_secret_envelope}. We first restrict attention to simple functions $Z_m=h_m(Y)$ with $m+1$ outcomes. Let $\cG$ with $|\cG|=1$ denote the \say{good} set such that $P_Z(\cG)=1-\delta$, and let $\cB_m$ denote the
\say{bad} set such that $P_Z(\cB_m)=\delta$, with
\[
\max_{z\in\cG}\ell(z)\le \min_{z\in\cB_m}\ell(z).
\]

Suppose first that $m=2$. Using tail events as in Theorem~\ref{thm:Gauss_mech_Gauss_secret_envelope}, the \say{worst} choice
$Z_2^*$ satisfies
\begin{align*}
    \bar \varepsilon_{Z_2^*} (\delta) = \log \frac{2}{\delta}.    
\end{align*}

Next, suppose $m=3$. Let the outcomes of $Z_3$ correspond to three events $\cE_i$ with $P_Y(\cE_i)=\delta_i$ for $i=1,2,3$, such that $\sum_{i=1}^3 \delta_i=\delta$. By Lemma~\ref{lemma:tail_worse}(i), we may take
$\cE_1=(-\infty,t_L)$ and $\cE_2=(t_R,\infty)$, where $t_L,t_R$ are chosen so that $P_Y(\cE_1)=\delta_1$ and
$P_Y(\cE_2)=\delta_2$. When $\delta$ is sufficiently small, Lemma~\ref{lemma:tail_worse}(iii) allows us to restrict our candidates for $\cE_3$ to an interval adjacent to one of the tails. Suppose $\cE_3=(t_C,t_R]$, where $t_C$ is selected so that $P_Y(t_C,t_R]=\delta_3$. By Lemma~\ref{lemma:interval_leakage_increasing_general}, we have
\begin{equation*}
    \ell(X \to \cE_3)
    \leq \ell\bigl(X \to (t_C,\infty)\bigr)
    = \ell\bigl(X \to \cE_3 \cup \cE_2\bigr)
    \leq \ell\bigl(X \to (t_R,\infty)\bigr)
    = \ell(X \to \cE_2).
\end{equation*}
Therefore, we obtain
\begin{align*}
    \bar \varepsilon_{Z_3} (\delta)
    &= \min \Big\{\ell(X \to \cE_1), \ell(X \to \cE_2), \ell(X \to \cE_3) \Big\}\\
    &= \min \Big\{\ell(X \to \cE_1), \ell(X \to \cE_3) \Big\}\\
    &\leq \min \Big\{\ell(X \to \cE_1), \ell(X \to \cE_3 \cup \cE_2) \Big\}\\
    &\leq \max_{\substack{\delta_1,\delta_2,\delta_3>0:\\ \delta_1+\delta_2+\delta_3=\delta}}
    \min \Big\{\log \frac{1}{\delta_1}, \log \frac{1}{\delta_2+\delta_3}\Big\}\\
    &= \log \frac{2}{\delta}.
\end{align*}
The other case $\cE_3=[t_L,t_C')$ is handled identically and cannot increase the PML $\delta$-quantile beyond $\log\frac{2}{\delta}$. Hence, no $Z_3$ can have a larger PML $\delta$-quantile than $Z_2^*$.

Repeating the same merging step iteratively shows that the worst-case PML $\delta$-quantile over all simple
$Z_m=h_m(Y)$ is already attained for $Z_2^*$, and hence
\begin{equation*}
    \bar{\varepsilon}_{h(Y)}(\delta) \leq \log\frac{2}{\delta},
\end{equation*}
for every simple $h$. The extension to an arbitrary measurable $h$ follows from Lemma~\ref{lemma:simple_to_measurable}.

\subsection{Proof of Lemma~\ref{lemma:interval_leakage_increasing_general}}
\label{ssec:lemma_interval_leakage_increasing_general_proof}
Fix $a>0$ and let 
\begin{align}
    \ell'(X \to (a,b)) &=  \frac{d}{db} \left(\log \, \Big( 2\Phi(\frac{b-a}{2 \sigma_N}) -1 \Big) - \log \Big(F_Y(b) - F_Y(a)\Big) \right)\nonumber\\
    &= \frac{\varphi(\frac{b-a}{2\sigma_N})}{\sigma_N \left(2\Phi(\frac{b-a}{2 \sigma_N}) -1\right)} - \frac{f_Y(b)}{F_Y(b) - F_Y(a)}. 
\end{align}
Our approach is similar to the proof of Lemma~\ref{lemma:interval_leakage_increasing}. In particular, we first show that $\ell(X \to (a,b))$ is increasing when $b$ is sufficiently close to $a$. This is stated in the form of the following lemma proved in Appendix~\ref{ssec:lemma_event_leakage_first_increase_general_proof}. 

\begin{lemma} 
\label{lemma:event_leakage_first_increase_general}
Suppose $a > M$. There exists $\xi_0 > 0$ such that the mapping $b \mapsto \ell(X \to (a, b))$ is strictly increasing on the interval $(a, a + \xi_0)$. 
\end{lemma}

Now, let $u(b) = \ell'(X \to (a,b)) = r(b) - s(b)$, where   
\begin{gather*}
    r(b) \coloneqq \frac{\frac{1}{\sigma_N} \varphi(\frac{b-a}{2\sigma_N})}{2\Phi(\frac{b-a}{2\sigma_N}) -1},\\
    s(b) \coloneqq \frac{f_Y(b)}{F_Y(b) - F_Y(a)}.
\end{gather*}
We calculate the derivatives of $r$ and $s$ and express them as differential equations: 
\begin{align*}
    r'(b) &= - \frac{\frac{(b-a)}{4\sigma_N^3} \varphi(\frac{b-a}{2\sigma_N}) \Big(2 \Phi(\frac{b-a}{2\sigma_N}) -1 \Big) + \frac{1}{\sigma_N^2} \varphi^2(\frac{b-a}{2\sigma_N})}{\left(2\Phi(\frac{b-a}{2\sigma_N}) -1\right)^2}\nonumber\\[0.6em]
    &= - \frac{(b-a)}{4\sigma_N^2} r(b) - r^2(b),\\[1em]
    s'(b) &= \frac{f'_Y(b) \big(F_Y(b) - F_Y(a)\big) - f_Y^2(b)}{\big(F_Y(b) - F_Y(a)\big)^2}\\
    &= \frac{f'_Y(b)}{F_Y(b) - F_Y(a)} - s^2(b). 
\end{align*}
The first term in the expression of $s'(b)$ can be written as 
\begin{align*}
    \frac{f'_Y(b)}{F_Y(b) - F_Y(a)} = \frac{f'_Y(b)}{f_Y(b)} s(b)= \frac{\big(\bE[X \mid Y=b] - b\big)}{\sigma_N^2} s(b),
\end{align*}
where the second equality is due to Tweedie's formula~\eqref{eq:tweedie}. So, we have
\begin{equation*}
   s'(b) =  \frac{\big(\bE[X \mid Y=b] - b\big)}{\sigma_N^2} s(b) - s^2(b). 
\end{equation*}
Thus, we can write $u'(b) = r'(b) - s'(b)$ as  
\begin{align*}
    u'(b) &= - \frac{(b-a)}{4\sigma_N^2} \big(u(b)+s(b)\big) - \big(u(b) + s(b)\big)^2 - \frac{\big(\bE[X \mid Y=b] - b\big)}{\sigma_N^2} s(b) + s^2(b),\\
    &= -u^2(b) - \frac{(b-a)}{4\sigma_N^2} u(b) - 2 u(b) s(b) + \Big(\frac{3b + a - 4\, \bE[X \mid Y=b]}{4\sigma_N^2}\Big) s(b).
\end{align*}

Now, suppose $b_0 \in (a,\infty)$ is a local optimum of $\ell(X \to (a,b))$, that is $u(b_0) = 0$. At this point, we would have 
\begin{equation*}
    u'(b_0) = \frac{3b_0 + a - 4\, \bE[X \mid Y=b_0]}{4\sigma_N^2} s(b_0).
\end{equation*}
Since $s(b) >0$ for all $b \in (a,\infty)$, the sign of $u'(b_0)$ is determined by the sign of $3b_0 + a - 4 \bE[X \mid Y=b_0]$. We write 
\begin{align*}
   3b_0 + a - 4 \bE[X \mid Y=b_0] &= 3b_0 + a - 4 \bE[X \mid Y=b_0] + 3a - 3a\\
   &= 3(b_0 - a) + 4 \big(a - \bE[X \mid Y=b_0] \big)\\
   &= 3(b_0 - a) \Big( 1 + \frac{4}{3} \cdot \frac{a - \bE[X \mid Y=b_0]}{b_0 - a} \Big)\\
   &= 3(b_0 - a) \Big( 1 + \frac{4}{3} \cdot \frac{a - \bE[X \mid Y=a] + \bE[X \mid Y=a] - \bE[X \mid Y=b_0]}{b_0 - a} \Big)\\
   &> 3(b_0 - a) \Big( 1 - \frac{4}{3} \cdot \frac{\bE[X \mid Y=b_0] - \bE[X \mid Y=a]}{b_0 - a} \Big),
\end{align*}
where the last inequality is because Tweedie's formula \eqref{eq:tweedie} states that $\bE[X \mid Y=a] - a< 0$ when $f'_Y(a) < 0$, which holds since $a > M$. Hence, having 
\begin{equation}
\label{eq:lipschitz_condition}
    \sup_{b \in (a,\infty)}\frac{\bE[X \mid Y=b] - \bE[X \mid Y=a]}{b - a} \leq  \frac{3}{4},
\end{equation}
is a sufficient condition for guaranteeing that $3b_0 + a - 4 \bE[X \mid Y=b_0] >0$, and in turn $u'(b_0) > 0$. Note that since $b \mapsto \bE[X \mid Y=b]$ is differentiable (this follows from Tweedie's formula~\eqref{eq:tweedie} and that $f_Y$ is real analytic), a sufficient condition for ensuring \eqref{eq:lipschitz_condition} is   
\begin{equation*}
    \sup_{b \in (a,\infty)} \; \frac{d}{db} \bE[X \mid Y=b] \leq \frac{3}{4}. 
\end{equation*}
Earlier, we had calculated $\frac{d}{db} \bE[X \mid Y=b]$ in \eqref{eq:derivative_posterior_mean}. Plugging in that expression, we get
\begin{align*}
    \sup_{b \in (a,\infty)} \; \frac{d}{db} \bE[X \mid Y=b] = \sup_{b \in (a,\infty)} \; \frac{\operatorname{Var}[X \mid Y=b]}{\sigma_N^2} \leq \sup_{b \in (-\infty,\infty)} \; \frac{\operatorname{Var}[X \mid Y=b]}{\sigma_N^2}. 
\end{align*}
We conclude that 
\begin{equation*}
    \sup_{y \in (-\infty,\infty)} \; \operatorname{Var}[X \mid Y=y] \leq \frac{3}{4} \sigma_N^2, 
\end{equation*} 
is a sufficient condition for $u'(b_0) >0$. 

Finally, observe that $u'(b_0) >0$ means that $u(b)$ cannot have any zeros. This is because $u(b)$ is positive for $b$ sufficiently close to $a$, but at any potential zeros of $u(b)$, it can only change sign from negative to positive, which is impossible due to the continuity of $u(b)$. (See also the proof of Lemma~\ref{lemma:interval_leakage_increasing} for a more detailed argument.) Therefore, $u(b) = \ell'(X \to (a,b)) >0$ for all $b > a$ which means that the leakage $\ell(X \to (a,b))$ is monotonically increasing and approaches its limit 
\begin{equation*}
    \lim_{b \to \infty} \, \ell(X \to (a,b)) = \log \frac{1}{P_Y(a,\infty)}. 
\end{equation*}

\subsection{Proof of Lemma~\ref{lemma:event_leakage_first_increase_general}}
\label{ssec:lemma_event_leakage_first_increase_general_proof}
Fix $a >M$. We show that $\frac{d}{d \xi} \ell(X \to (a, a+ \xi)) > 0$ for sufficiently small $\xi >0$. This derivative can be written as 
\begin{align*}
    \frac{d}{d \xi} \; \ell(X \to (a, a+ \xi)) &= \frac{d}{d \xi}  \; \log \, \frac{2\Phi(\frac{\xi}{2 \sigma_N}) -1}{F_Y(a+\xi) - F_Y(a)}\\
    &= \frac{d}{d \xi} \Big(\log \big( 2\Phi(\frac{\xi}{2 \sigma_N}) -1 \big) - \log \big( F_Y(a+\xi) - F_Y(a) \big)\Big) \\
    &= \frac{\frac{1}{\sigma_N} \varphi(\frac{\xi}{2 \sigma_N})}{2\Phi(\frac{\xi}{2 \sigma_N}) -1} - \frac{f_Y(a + \xi)}{F_Y(a+\xi) - F_Y(a)}\\
    &= \frac{\frac{1}{\sigma_N} \varphi(\frac{\xi}{2 \sigma_N})\Big(F_Y(a+\xi) - F_Y(a)\Big) - f_Y(a + \xi) \Big(2\Phi(\frac{\xi}{2 \sigma_N}) -1\Big)}{\Big(2\Phi(\frac{\xi}{2 \sigma_N}) -1\Big) \Big(F_Y(a+\xi) - F_Y(a)\Big)}.
\end{align*}
The denominator is clearly positive, so we focus on showing that the numerator is positive. For this, we use the Taylor expansions 
\begin{gather*}
     \varphi(\frac{\xi}{2\sigma_N}) = \varphi(0) \left(1 - \frac{1}{8 \sigma_N^2} \xi^2 \right) + O(\xi^4), \\
     F_Y(a+\xi) - F_Y(a) = f_Y(a) \xi + \frac{1}{2} f'_Y(a) \xi^2 + O(\xi^3),\\
     f_Y(a + \xi) = f_Y(a) + f'_Y(a) \xi + O(\xi^2), \\
     2\Phi(\frac{\xi}{2 \sigma_N}) -1 = \frac{\varphi(0)}{\sigma_N}\xi-
        \frac{\varphi(0)}{24\,\sigma_N^{3}}\,\xi^{3}
        + O(\xi^{5}).
\end{gather*}
Hence, we have
\begin{align}
    \frac{1}{\sigma_N} \varphi(\frac{\xi}{2 \sigma_N})&\Big(F_Y(a+\xi) - F_Y(a)\Big) - f_Y(a + \xi) \Big(2\Phi(\frac{\xi}{2 \sigma_N}) -1\Big) \nonumber\\
    &= \frac{\varphi(0)}{\sigma_N} f_Y(a) \xi + \frac{\varphi(0)}{2\sigma_N} f'_Y(a) \xi^2 - \frac{\varphi(0)}{\sigma_N}  f_Y(a) \xi -\frac{\varphi(0)}{\sigma_N} f'_Y(a) \xi^2 +  O(\xi^3) \nonumber \\
    &= - \frac{\varphi(0)}{2\sigma_N} f'_Y(a) \xi^2 +  O(\xi^3). \label{eq:sub-Gaussian_derivative_numerator} 
\end{align}
By assumption, $a > M$, so $f'_Y(a) < 0$. It follows that \eqref{eq:sub-Gaussian_derivative_numerator} is positive for sufficiently small $\xi >0$.

\end{document}